\documentclass{amsart}%
\usepackage{amsmath}
\usepackage{amssymb}
\usepackage{graphicx}
\usepackage{amscd}%
\usepackage{amsfonts}
\newtheorem{theorem}{Theorem}
\theoremstyle{plain}

\newtheorem{corollary}{Corollary}

\newtheorem{definition}{Definition}

\newtheorem{proposition}{Proposition}
\newtheorem{remark}{Remark}

\numberwithin{equation}{section}

\begin{document}
\title[Quantum Feedback Control]{Towards the theory of control in observable quantum systems}
\author{V. P. Belavkin}
\address{Moscow Institute of Electronics and Mathematics, 3/12 B. Vuzovski Per. Moscow
109028, USSR}
\date{April 8, 1981}

\begin{abstract}
The fundamental mathematical definitions of the controlled feedback Markov
dynamics of quantum-mechanical systems are introduced with regard to the
dynamical reduction and filtering of quantum states in the course of quantum
measurement in either discrete or continuous real time. \ The concept of
sufficient coordinates for the description of \textit{a posteriori} quantum
states in a given class is introduced and it is proved that they form a
classical Markov process with values in either state operators or state vector
space. \ The general problem of optimal control of a quantum-mechanical system
is discussed and the corresponding Bellman equation in the space of sufficient
coordinates is derived. \ The results are illustrated in the example of
control of the semigroup dynamics of a quantum system that is instantaneously
observed at discrete times and evolves between measurement times according to
the Schr\H{o}dinger equation.

\end{abstract}
\maketitle

\section{Introduction}

The encouraging outlook for the application of coherent quantum optics
(lasers) for communications and control has been recently stimulated by the
steadily growing demands for greater accuracy of observation and monitoring,
particularly under the \textquotedblleft extreme\textquotedblright\ conditions
of very faint signals at extremely great (astronomical) distances. On the
other hand, instances of the successful exploitation of mathematical methods
from information and control theory for the investigation of many physical
phenomena in the microscopic world have also stimulated interest in the
theoretical study, using general cybernetic principles, of the possibilities
of dynamical systems described at the quantum-mechanical level \cite{Sam72}%
\cite{Kra73}\cite{Kra74}\cite{Be74}\cite{Be78b}. \ It has been shown in
\cite{But_Sam79} that it is natural to regard many physical problems as
control problems for distributed systems described by standard
quantum-mechanical equations. \ In particular, the possibility of the
transition of a physical system from one microscopic state to another can be
investigated \cite{But_Sam80} by the methods of the theory of controllability
on Lie groups generated by the Schr\H{o}dinger equation with a controlled Hamiltonian.

General problems in the theory of quantum dynamical systems with observation,
control and feedback channels can be handled on the basis of the recent
development \cite{Be78a} of an operational theory of open-loop quantum
systems, for which the mathematical formalism was set down in \cite{Dav77}
\cite{Ing75}. The investigation, undertaken in \cite{Be78}, of the dynamical
observation and feedback control optimization problems for such systems has
provided a means for solving these problems in the case of linear Markov
systems of the boson type, in particular for a controllable and observable
quantum oscillator \cite{Be78b}. \ This work was based on the multistage
quantum-statistical decision theory originally described in \cite{Be79}%
\cite{Be80a} for the problems of the optimal dynamical measurement and control
of classical (i.e., commutative) Markov processes with quantum observation channels.

In the present article we describe a simplified problem of optimal feedback
control of quantum dynamical systems which does not involve
quantum-statistical decision theory. \ Here the observable subsystem at the
output of the observable channel is regarded as classical and amendable to
description at the macroscopic level, whereas the controlled entity remains a
quantum dynamical system. \ In other words, we assume here, in contrast with
\cite{Be78}\cite{Be79}\cite{Be80a}, that the \textquotedblleft
instrument\textquotedblright\ at the output of the quantum-mechanical system
is given, rather than to be optimized, and it is required only to find the
optimal macroscopic feedback for a given performance criteria. The results
obtained in this setting are special in relation to \cite{Be78b}%
\cite{Be78a}\cite{Be78} as they correspond to the semiclassical case of
commutativity of the algebra of output observables. \ They nonetheless deserve
special consideration both from the methodological and from the practical
point of view when the observation channels are given and cannot be optimized
for the optimal feedback control purpose.

\section{\label{Section2}Controllable quantum dynamical systems with
observation}

Here we introduce the mathematical concept of controllable quantum system with
observation channel on the basis of the operational theory on open-loop
physical systems and quantum processes \cite{Be78a}\cite{Dav77}\cite{Ing75}.
\ Such systems are open by the definition, and the necessary concepts borrowed
from the algebraic theory of open quantum systems are described in the Appendix.

Let $\mathcal{H}$ be the Hilbert space of representation of a certain
quantum-mechanical system regarded as an observable and controllable system
and let $\mathfrak{A}$ be the von Neumann algebra of admissible physical
quantities $Q\in\mathfrak{A}$ which is generated (see Appendix 1) by the
dynamical variables of this system, acting as operators in $\mathcal{H}$.
\ The pair $\left\{  \mathcal{H},\mathfrak{A}\right\}  $ plays the role of a
measurable space $\left\{  X,\mathcal{A}\right\}  $ representing \cite{Kol74}
the corresponding classical dynamical system in the phase space $X$ of it's
point states, endowed with the Borel $\sigma$-algebra $\mathcal{A}$ of
admissible events $A\in\mathcal{A}$. \ The simple systems normally treated in
traditional texts on quantum mechanics, for example \cite{Lan_Lif63},
correspond to the algebras $\mathfrak{A}=\mathcal{B}\left(  \mathcal{H}%
\right)  $ of all bounded operators in $\mathcal{H}$, but the models that
emerge from quantum field theory and statistical mechanics \cite{Emc72} are
described by the more general algebras $\mathfrak{A}$.

Normal states of the quantum-mechanical system at every time $t\in\mathbb{R} $
are determined by the linear functionals $\varrho_{t}:Q\mapsto\left\langle
\rho_{t},Q\right\rangle $ of the quantum-mechanical expectations $\left\langle
Q\right\rangle _{t}=\left\langle \rho_{t},Q\right\rangle $ of all the physical
quantities $Q\in\mathfrak{A}$ for this system, and are described by the
densities $\rho_{t}$ as the positive elements associated with von Neumann
algebra $\mathfrak{A}$ (see Appendix 2). In the case of semifinite algebras
\cite{Dix69}, as in the simple case $\mathfrak{A}=\mathcal{B}\left(
\mathcal{H}\right)  $, the states $\varrho_{t} $ are usually represented by
the trace one operators $\rho_{t}$ in $\mathfrak{A}$ (or affiliated with
$\mathfrak{A}$) as%
\begin{equation}
\varrho_{t}\left(  Q\right)  =\mathrm{tr}\left\{  \rho_{t}Q\right\}
\equiv\left\langle \rho_{t},Q\right\rangle .\label{1.1}%
\end{equation}

The Master evolution $t\mapsto\varrho_{t}$ of a quantum-mechanical system
controlled on each time interval $[t,t+\tau)$ by a segment $u_{t}^{\tau
}=\left\{  u\left(  r\right)  :r\in\lbrack t,t+\tau)\right\}  $ of
certain\ parameters $u\left(  t\right)  $\ is usually described by linear
unital completely positive transformations of the states
\begin{equation}
\varrho_{t}\in\mathfrak{A}_{\star}\mapsto\varrho_{t}\mathrm{M}_{t}^{\tau
}\left(  u_{t}^{\tau}\right)  \equiv\varrho_{t+\tau}\left(  u_{t}^{\tau
}\right)  \in\mathfrak{A}_{\star}\text{ \ \ \ \ }\forall t\in\mathbb{R}%
,\tau>0.\label{1.2}%
\end{equation}
They are determined by the composition of the functionals $\varrho
_{t}:\mathfrak{A}\mapsto\mathbb{C}$ with controlled transfer operators
$\mathrm{M}_{t}^{\tau}\left(  u_{t}^{\tau}\right)  $ as the maps
$\mathfrak{A}\mapsto\mathfrak{A}$ defined in the Appendix 3. \ The Markov
family $\left\{  \mathrm{M}_{t}^{\tau}\right\}  _{t\in\mathbb{R},\text{ }%
\tau>0}$ of these maps must satisfy the consistency condition
\begin{equation}
\mathrm{M}_{r}^{\tau}\left(  u_{r}^{\tau}\right)  \mathrm{M}_{r+\tau}%
^{\tau^{\prime}}\left(  u_{r+\tau}^{\tau^{\prime}}\right)  =\mathrm{M}%
_{r}^{\tau+\tau^{\prime}}\left(  u_{r}^{\tau+\tau^{\prime}}\right)  \text{
\ \ \ \ }\forall r\in\mathbb{R}\text{, \ \ \ \ }\tau,\tau^{\prime
}>0,\label{1.3}%
\end{equation}
as quantum analog of the controlled Chapman-Kolmogorov equation. Here
$u_{r}^{\tau}=\left\{  u\left(  r+\tau^{\prime}\right)  \right\}
_{\tau^{\prime}<\tau}$ is a segment of an admissible control function
$u\left(  t\right)  \in U\left(  t\right)  $ of length $\tau>0$ and
$u_{r}^{\tau+\tau^{\prime}}$ denotes the composition $\left(  u_{r}^{\tau
},u_{r+\tau}^{\tau^{\prime}}\right)  $ of segments $u_{r}^{\tau}$ and
$u_{r+\tau}^{\tau^{\prime}}$. Note that for stationary (time shift-invariant)
systems, where $U\left(  t\right)  =U$, (\ref{1.3}) specifies the cocycle
conditions for transfer operators $\mathrm{M}^{\tau}\left(  u^{\tau}\right)
=\mathrm{M}_{r}^{\tau}\left(  u_{r}^{\tau}\right)  $ independent of $r$ with
respect to the shift $u^{\tau}\mapsto u_{r}^{\tau}$ given by $u_{r}\left(
t\right)  =u\left(  t+r\right)  $.

In the spatial case of transfer operators $\mathrm{M}_{t}^{\tau}\left(
u_{t}^{\tau}\right)  $ specified by controllable propagators $T\left(
u_{t}^{\tau}\right)  :\mathcal{H}\mapsto\mathcal{H}$ satisfying the condition
corresponding to (\ref{1.2}),
\[
T_{r+\tau}^{\tau^{\prime}}\left(  u_{r+\tau}^{\tau^{\prime}}\right)
T_{r}^{\tau}\left(  u_{r}^{\tau}\right)  =T_{r}^{\tau+\tau^{\prime}}\left(
u_{r}^{\tau+\tau^{\prime}}\right)  ,
\]
the dynamics $\varrho_{t}\mapsto\varrho_{t+\tau}\left(  u_{t}^{\tau}\right)  $
for the vector states $\left\langle \rho_{t},Q\right\rangle =\left\langle
\psi_{t}|Q\psi_{t}\right\rangle $ is described by the state-vector
transformations
\[
\psi_{t}\in\mathcal{H}\mapsto T_{t}^{\tau}\left(  u_{t}^{\tau}\right)
\psi_{t}\equiv\psi_{t+\tau}\left(  u_{t}^{\tau}\right)  \in\mathcal{H}\text{
\ \ \ \ }\forall t\in\mathbb{R},\tau>0.
\]
Such transformations can be obtained, for example, as the fundamental
solutions of the time-dependant Schr\H{o}dinger equation with a perturbing
force $u\left(  t\right)  $, for which the isometric operators $T_{t}^{\tau
}\left(  u_{t}^{\tau}\right)  $ are unitary, and $\mathrm{M}_{t}^{\tau}\left(
u_{t}^{\tau}\right)  $ are the Heisenberg transformations%
\[
\mathrm{M}_{t}^{\tau}\left(  u_{t}^{\tau}\right)  Q=T_{t}^{\tau}\left(
u_{t}^{\tau}\right)  ^{\dagger}QT\left(  u_{t}^{\tau}\right)  .
\]

The open-loop input-output quantum dynamical systems of the kind specified
below as controllable systems with observation cannot, as a rule, be described
in terms of propagators $T_{t}^{\tau}\left(  u_{t}^{\tau}\right)  $, because
the measurements induce reductions and decoherence of quantum states which are
described by more general transformations. In order to define such systems as
quantum dynamical objects with the classical inputs $u\left(  t\right)  $ and
outputs $v\left(  t\right)  \in V\left(  t\right)  $, we shall fix a pair of
two-parameter families $\left\{  U_{t}^{\tau}\right\}  ,\left\{  V_{t}^{\tau
}\right\}  ,t\in\mathbb{R},\tau>0$ of the Cartesian product subsets
$U_{t}^{\tau}\subseteq\prod\limits_{\tau^{\prime}<\tau}U\left(  t+\tau
^{\prime}\right)  $, $V_{t}^{\tau}=\prod\limits_{\tau^{\prime}<\tau}V\left(
t+\tau^{\prime}\right)  $, satisfying the consistency condition
\begin{equation}
U_{r}^{\tau}\times U_{r+\tau}^{\tau^{\prime}}=U_{r}^{\tau+\tau^{\prime}%
}\text{, \ \ \ \ }V_{r}^{\tau}\times V_{r+\tau}^{\tau^{\prime}}=V_{r}%
^{\tau+\tau^{\prime}}.\label{1.4}%
\end{equation}
The elements $u_{t}^{\tau}\in U_{t}^{\tau},v_{t}^{\tau}\in V_{t}^{\tau}$ are
called the admissible segments of signals, and their Cartesian compositions
should also be admissible in order to make sense, for example, of (\ref{1.3}).
The sets $U_{r}^{\tau}$ are usually endowed with Hausdorff product topologies,
and the sets $V_{r}^{\tau}$ with Borel product $\sigma$-algebras, making them
consistent with (\ref{1.4}). Note that the continuous signals $u\left(
t\right)  $ and $v\left(  t\right)  $ on $t\in\mathbb{R}$ do not satisfy the
conditions (\ref{1.4}). In the time shift-invariant case $U_{r}^{\tau}$ and
$V_{r}^{\tau}$ are given by the shift $t\mapsto t+r$ of the initial sets
$U^{\tau}=U_{0}^{\tau}$ and $V^{\tau}=V_{0}^{\tau}$.

\begin{definition}
A controllable quantum dynamical system with observation is described by a
family $\left\{  \Pi_{t}^{\tau}\right\}  _{t\in\mathbb{R}\text{, }\tau>0}$ of
controllable transfer-operator measures $\Pi_{t}^{\tau}\left(  u_{t}^{\tau
},dv_{t}^{\tau}\right)  :\mathfrak{A}\mapsto\mathfrak{A}$ (see Appendix 4)
defined on the spaces $U_{t}^{\tau}\ni u_{t}^{\tau},V_{t}^{\tau}\supseteq
dv_{t}^{\tau}$ and satisfying the condition
\begin{equation}
\Pi_{r}^{\tau}\left(  u_{r}^{\tau},dv_{r}^{\tau}\right)  \Pi_{r+\tau}%
^{\tau^{\prime}}\left(  u_{r+\tau}^{\tau^{\prime}},dv_{r+\tau}^{\tau^{\prime}%
}\right)  =\Pi_{r}^{\tau+\tau^{\prime}}\left(  u_{r}^{\tau+\tau^{\prime}%
},dv_{r}^{\tau+\tau^{\prime}}\right) \label{1.5}%
\end{equation}
for any $r\in\mathbb{R},$ $\tau,\tau^{\prime}>0$, where $u_{r}^{\tau
+\tau^{\prime}}=\left(  u_{r}^{\tau},u_{r+\tau}^{\tau^{\prime}}\right)  $,
$dv_{r}^{\tau+\tau^{\prime}}=dv_{r}^{\tau}\times dv_{r+\tau}^{\tau^{\prime}}.$
\end{definition}

The superoperators $\Pi_{t}^{\tau}\left(  u_{t}^{\tau},dv_{t}^{\tau}\right)  $
are assumed to be continuous in $u_{t}^{\tau}$, $\sigma$-additive with respect
to $dv_{t}^{\tau}$ (in the strong operator sense) and in the time
shift-invariant case not to depend explicitly on $t$ [the index $t$ in the
condition (\ref{1.5}) specifying the cocycle dependence on $\tau$ can now be omitted].

Due to positivity $Q\in\mathfrak{A}_{+}\mapsto\Pi_{t}^{\tau}\left(
u_{t}^{\tau},dv_{t}^{\tau}\right)  Q\in\mathfrak{A}_{+}$ and the normalization
condition $\Pi_{t}^{\tau}\left(  u_{t}^{\tau},V_{t}^{\tau}\right)
=\mathrm{M}_{t}^{\tau}\left(  u_{t}^{\tau}\right)  $, the mappings $\Pi
_{t}^{\tau}$ determine, for a given instantaneous state $\varrho_{t}%
\in\mathfrak{A}_{\star}$ and control function $u_{t}^{\tau}$, the future
$\left(  \tau>0\right)  $ states\bigskip%
\begin{equation}
\varrho_{t+\tau}\left(  u_{t}^{\tau},dv_{t}^{\tau}\right)  =\varrho_{t}\Pi
_{t}^{\tau}\left(  u_{t}^{\tau},dv_{t}^{\tau}\right) \label{1.6}%
\end{equation}
of the quantum-mechanical process, normalized to the probabilities
\begin{equation}
\pi_{t}^{\tau}\left(  u_{t}^{\tau},dv_{t}^{\tau}\right)  =\left\langle
\rho_{t},\Pi_{t}^{\tau}\left(  u_{t}^{\tau},dv_{t}^{\tau}\right)
I\right\rangle \equiv\varrho\Pi_{t}^{\tau}\left(  u_{t}^{\tau},dv_{t}^{\tau
}\right)  I\label{1.7}%
\end{equation}
of the events $v_{t}^{\tau}\in dv_{t}^{\tau}$. \ The ratio of (\ref{1.6}) to
(\ref{1.7}) determines conditional states, depending in general non-linearly
on $\varrho=\varrho_{t}$:
\begin{equation}
\varrho_{t}^{\tau}=\frac{\varrho\Pi_{t}^{\tau}\left(  u_{t}^{\tau}%
,dv_{t}^{\tau}\right)  }{\varrho\Pi_{t}^{\tau}\left(  u_{t}^{\tau}%
,dv_{t}^{\tau}\right)  I}.\label{1.8}%
\end{equation}
Note that the normalized conditional states are well-defined only for the
measurable events $dv_{t}^{\tau}\subseteq V_{t}^{\tau}$ of non-zero
probability (\ref{1.7}) with which the system transfers from the state
$\varrho_{t}=\varrho$ as a result of the control action $u_{t}^{\tau}$ and the
observation $dv_{t}^{\tau}$ on an interval of length $\tau$. \ If the event
$dv_{t}^{\tau}=V_{t}^{\tau}$ is certain, $\pi_{t}^{\tau}\left(  u_{t}^{\tau
},dv_{t}^{\tau}\right)  =1$, the states (\ref{1.8}) are unconditional such
that $\varrho_{t}^{\tau}$ coincide with \textit{a priori }states
$\varrho_{t+\tau}\left(  u_{t}^{\tau}\right)  $, whose controlled evolution is
linear. \ Such evolution is described by expression (\ref{1.2}), in which
$\mathrm{M}_{t}^{\tau}\left(  u_{t}^{\tau}\right)  =\Pi_{t}^{\tau}\left(
u_{t}^{\tau},V_{t}^{\tau}\right)  $ denotes controllable transforms of the
open-loop quantum-mechanical system, corresponding to the absence of
observation. However in general\ the process of precise measurement of the
output signal $v_{t}^{\tau}$ on an interval of length $\tau>0$ takes the
quantum system from \textit{a priori} state $\varrho=\varrho_{t}$ to the
\textit{a posteriori} state $\varrho_{t}^{\tau}=\varrho\mathrm{M}_{\rho
,t}^{\tau}\left(  u_{t}^{\tau},v_{t}^{\tau}\right)  $, where the quasilinear
mapping $\varrho\mapsto\varrho\mathrm{M}_{\varrho,t}^{\tau}\left(  u_{t}%
^{\tau},v_{t}^{\tau}\right)  $ is given by expression (\ref{1.8}) in the limit
$dv_{t}^{\tau}\downarrow\left\{  v_{t}^{\tau}\right\}  $ almost everywhere
with respect to the measure (\ref{1.7}). \ For example, let the measures
$\Pi_{t}^{\tau}$ have the density functions
\begin{equation}
\Pi_{t}^{\tau}\left(  u_{t}^{\tau},dv_{t}^{\tau}\right)  =\int\limits_{dv_{t}%
^{\tau}}\mathrm{P}_{t}^{\tau}\left(  u_{t}^{\tau},v_{t}^{\tau}\right)  \mu
_{t}^{\tau}\left(  u_{t}^{\tau},dv_{t}^{\tau}\right)  ,\label{1.9}%
\end{equation}
where $\mathrm{P}_{t}^{\tau}\left(  u_{t}^{\tau},v_{t}^{\tau}\right)
:\mathfrak{A}\mapsto\mathfrak{A}$ denotes completely positive superoperators
[see the Appendix, (A.4)], say, of the form (\ref{2.10}), continuous with
respect to $u_{t}^{\tau}$ and integrable with respect to $v_{t}^{\tau}$ in the
strong operator sense with respect to specified positive measures $\mu
_{t}^{\tau}$ on $V_{t}^{\tau}$. \ Then the \textit{a posteriori} transfer
operators $\mathrm{M}_{\rho,t}^{\tau}\left(  u_{t}^{\tau},v_{t}^{\tau}\right)
$ coincide, up to normalization, with $\mathrm{P}_{t}^{\tau}\left(
u_{t}^{\tau},v_{t}^{\tau}\right)  :$%
\begin{equation}
\mathrm{M}_{\rho,t}^{\tau}\left(  u_{t}^{\tau},v_{t}^{\tau}\right)
=\frac{\mathrm{P}_{t}^{\tau}\left(  u_{t}^{\tau},v_{t}^{\tau}\right)
}{\left\langle \rho,\mathrm{P}_{t}^{\tau}\left(  u_{t}^{\tau},v_{t}^{\tau
}\right)  I\right\rangle },\label{1.10}%
\end{equation}
where the ratio is defined for those $u_{t}^{\tau}\in U_{t}^{\tau},$
$v_{t}^{\tau}\in V_{t}^{\tau}$ for which the densities
\begin{equation}
p_{t}^{\tau}\left(  u_{t}^{\tau},v_{t}^{\tau}\right)  =\left\langle
\rho,\mathrm{P}_{t}^{\tau}\left(  u_{t}^{\tau},v_{t}^{\tau}\right)
I\right\rangle \equiv\varrho\mathrm{P}_{t}^{\tau}\left(  u_{t}^{\tau}%
,v_{t}^{\tau}\right)  I\label{1.11}%
\end{equation}
of the probability measure (\ref{1.7}) with respect to $\mu_{t}^{\tau}$ are non-vanishing.

The following theorem states that the \textit{a posteriori} mapping
(\ref{1.10}) in fact determines the state-valued classical Markov process,
which was introduced in the classical case by Stratonovich in \cite{Str60} (He
called this probability measure-valued process secondary, or conditional
(\textit{a posteriori}) Markov process).

\begin{theorem}
\label{Theorem1}The family $\left\{  \mathrm{M}_{\rho,t}^{\tau}\right\}  $ of
a posteriori transfer operators $\mathrm{M}_{\rho,t}^{\tau}\left(  u_{t}%
^{\tau},v_{t}^{\tau}\right)  $ satisfies, with respect to the operator
composition, the consistency condition
\begin{equation}
\mathrm{M}_{\rho,r}^{\tau}\left(  u_{r}^{\tau},v_{r}^{\tau}\right)
\mathrm{M}_{\rho^{\prime},r+\tau}^{\tau^{\prime}}\left(  u_{r+\tau}%
^{\tau^{\prime}},v_{r+\tau}^{\tau^{\prime}}\right)  =\mathrm{M}_{\rho,r}%
^{\tau+\tau^{\prime}}\left(  u_{r}^{\tau+\tau^{\prime}},v_{r}^{\tau
+\tau^{\prime}}\right) \label{1.12}%
\end{equation}
almost everywhere under the measure (\ref{1.7}), where $\rho^{\prime}$ is the
density of $\varrho^{\prime}=\varrho\mathrm{M}_{\rho,r}^{\tau^{\prime}}\left(
u_{r}^{\tau},v_{r}^{\tau}\right)  $.
\end{theorem}

\begin{proof}
It is required to verify the property (\ref{1.12}) for conditional mappings
(\ref{1.8}), for which it follows at once from the definition and (\ref{1.5}).
Then it applies also in the single point limit $dv_{t}^{\tau}\downarrow
\left\{  v_{t}^{\tau}\right\}  $. \ In the case (\ref{1.9}) the condition
(\ref{1.12}) is simply verified by computing the product (\ref{1.12}) of the
\textit{a posteriori} transfer operators (\ref{1.10}); for this purpose it is
necessary to invoke the corresponding condition
\begin{equation}
\mathrm{P}_{r}^{\tau}\left(  u_{r}^{\tau},v_{r}^{\tau}\right)  \mathrm{P}%
_{r+\tau}^{\tau^{\prime}}\left(  u_{r+\tau}^{\tau^{\prime}},v_{r+\tau}%
^{\tau^{\prime}}\right)  =\mathrm{P}_{r}^{\tau+\tau^{\prime}}\left(
u_{r}^{\tau+\tau^{\prime}},v_{r}^{\tau+\tau^{\prime}}\right)  .\label{1.13}%
\end{equation}
It is sufficient to require this composition condition for $V_{t}^{\tau
+\tau^{\prime}}$ almost everywhere $\left(  \operatorname{mod}\mu_{t}^{\tau
}\right)  $ and this will guarantees the satisfaction of condition (\ref{1.4})
if
\[
\mu_{r}^{\tau}\left(  u_{r}^{\tau},dv_{r}^{\tau}\right)  \mu_{r+\tau}%
^{\tau^{\prime}}\left(  u_{r+\tau}^{\tau^{\prime}},dv_{r+\tau}^{\tau^{\prime}%
}\right)  =\mu\left(  u_{r}^{\tau+\tau^{\prime}},dv_{r}^{\tau+\tau^{\prime}%
}\right)  .
\]

\end{proof}

\begin{remark}
If the superoperator densities $\mathrm{P}_{t}^{\tau}$ of the transition
measures (\ref{1.9}) preserve unity: $\mathrm{P}_{t}^{\tau}\left(  u_{t}%
^{\tau},v_{t}^{\tau}\right)  I=p_{t}^{\tau}\left(  u_{t}^{\tau},v_{t}^{\tau
}\right)  I$, the ratio (\ref{1.10}) determines $\rho$-independent transfer
operators $\mathrm{M}_{t}^{\tau}\left(  u_{t}^{\tau},v_{t}^{\tau}\right)  $
describing the controllable quantum dynamics of a system with two inputs $u$
and $v$. The second is an observable stochastic process with probability
measures $\pi_{t}^{\tau}\left(  u_{t}^{\tau},v_{t}^{\tau}\right)  =p_{t}%
^{\tau}\left(  u_{t}^{\tau},v_{t}^{\tau}\right)  \mu_{t}^{\tau}\left(
u_{t}^{\tau},v_{t}^{\tau}\right)  $ independent of the state of the system.
\ The a posteriori mappings (\ref{1.8}) in this case are linear, $\varrho
_{t}^{\tau}=\varrho\mathrm{M}_{t}^{\tau}\left(  u_{t}^{\tau},v_{t}^{\tau
}\right)  $, almost everywhere under the measure $\pi_{t}^{\tau}.$
\end{remark}

\section{\label{Section3}Sufficient coordinates of quantum-mechanical systems}

The description of the dynamics of simple closed-loop quantum-mechanical
systems for a certain class of initial states is known to be often reducible
to the determination of the time evolution of certain coordinates, the role of
which can be taken, for example, by vectors $\psi\in\mathcal{H}$, if only
vector initial states are considered. \ Some aspects of the controllability of
closed-loop quantum-mechanical systems described by a sufficient coordinate
$\psi_{t}\in\mathcal{H}$, satisfying the controlled Schr\H{o}dinger equation
have been recently investigated in \cite{But_Sam80}.

The concept of sufficient coordinates, which is introduced below for general
controllable quantum dynamical systems with observation and is intimately
related to the classical notion of sufficient statistics \cite{Str60}, plays
an even greater role for quantum control theory than the analogous concept in
stochastic control theory, because it permits control problems for
quantum-mechanical systems to be reduced to classical control problems with
localized or distributed parameters.

\begin{definition}
Let $X$ be a measurable space\footnote{For all practical purpoces it is always
sufficient to assume that $X$ is a standard Borel space, i.e. a complete
seperable metric space, also known as a Polish space (for example,
$\mathbb{R}^{n},\mathbb{C}^{n},$ or any countable set).}, and let $\left\{
\varrho_{x,t}\right\}  _{x\in X,t\in\mathbb{R}}$ be a family of states given,
for every $t\in\mathbb{R}$, by a measurable mapping $x\mapsto\varrho_{x,t}$ of
the space $X$ into the space of states $\varrho_{x,t}\in\mathfrak{A}_{\star}$
of a quantum-mechanical system at time $t$ such that the controlled evolution
(\ref{1.5}) of the system during an observation leaves this family up to a
normalization $\pi_{x,t}^{\tau}$ invariant:
\begin{equation}
\varrho_{x,t}\Pi_{t}^{\tau}\left(  u_{t}^{\tau},dv_{t}^{\tau}\right)
=\pi_{x,t}^{\tau}\left(  u_{t}^{\tau},dv_{t}^{\tau}\right)  \varrho
_{f_{x,t}^{\tau}\left(  u_{t}^{\tau},v_{t}^{\tau}\right)  ,t+\tau}\label{2.1}%
\end{equation}
Then $x\in X$ is called a sufficient coordinate for $\left\{  \varrho
_{x,t}\right\}  $, the controlled stochastic evolution of which $x\in X\mapsto
f_{x,t}^{\tau}\left(  u_{t}^{\tau},v_{t}^{\tau}\right)  \equiv x_{t}^{\tau}\in
X$ is described by the mappings $f_{x,t}^{\tau}:U_{t}^{\tau}\times V_{t}%
^{\tau}\rightarrow X$, continuous with respect to $u_{t}^{\tau}\in U_{t}%
^{\tau}$ and measurable with respect to $v_{t}^{\tau}\in V_{t}^{\tau}$ almost
everywhere under the measure
\[
\pi_{x,t}^{\tau}\left(  u_{t}^{\tau},dv_{t}^{\tau}\right)  =\left\langle
\rho_{x},\Pi_{t}^{\tau}\left(  u_{t}^{\tau},dv_{t}^{\tau}\right)
I\right\rangle .
\]

\end{definition}

Proceeding from (\ref{2.1}) taken in the limit $dv_{t}^{\tau}\searrow\left\{
v_{t}^{\tau}\right\}  $, we note that the density operators $x=\rho$ form the
sufficient coordinate space for the family of all normal states $\varrho$ on
$\mathfrak{A}$. It is given by the \textit{a posteriori} mapping $f_{\rho
,t}^{\tau}\left(  u_{t}^{\tau},v_{t}^{\tau}\right)  =\mathrm{M}_{\rho,t}%
^{\tau}\left(  u_{t}^{\tau},v_{t}^{\tau}\right)  _{\star}\rho$, provided only
[as in the case (\ref{1.9})] that there exists the derivative
\begin{equation}
\mathrm{M}_{\rho,t}^{\tau}\left(  u_{t}^{\tau},v_{t}^{\tau}\right)  =\frac
{\Pi_{t}^{\tau}\left(  u_{t}^{\tau},dv_{t}^{\tau}\right)  }{\pi_{\rho,t}%
^{\tau}\left(  u_{t}^{\tau},dv_{t}^{\tau}\right)  }.\label{2.2}%
\end{equation}
as the limit $dv_{t}^{\tau}\searrow\left\{  v_{t}^{\tau}\right\}  $.

\begin{theorem}
\label{Theorem2}The mappings $f_{x,t}^{\tau}$ in (\ref{2.1}) define a
sufficient statistics $x_{t}^{\tau}=f_{x,t}^{\tau}\left(  u_{t}^{\tau}%
,v_{t}^{\tau}\right)  $ for $\varrho_{t}\in\left\{  \varrho_{x,t}\right\}
_{x\in X}$ in the sense that the a posteriori states $\varrho_{t}^{\tau
}=\varrho\mathrm{M}_{\rho,t}^{\tau}\left(  u_{t}^{\tau},v_{t}^{\tau}\right)  $
are determined for such $\varrho=\varrho_{x,t}$ as $\varrho_{t}^{\tau}%
=\varrho_{x_{t}^{\tau},t+\tau}\forall\tau>0$. Moreover, the transition
probabilities
\begin{equation}
\pi_{x,t}^{\tau}\left(  u_{t}^{\tau},dx^{\prime}\right)  =\left\langle
\rho_{x},\Pi_{x,t}^{\tau}\left(  u_{t}^{\tau},dx^{\prime}\right)
\right\rangle ,\label{2.3}%
\end{equation}
from $x_{t}=x$ into $dx^{\prime}\ni x_{t}^{\tau}$, defined by%
\[
\Pi_{x,t}^{\tau}\left(  u_{t}^{\tau},dx^{\prime}\right)  =\Pi_{t}^{\tau
}\left(  u_{t}^{\tau},f_{x,t}^{-1}\left(  u_{t}^{\tau},dx^{\prime}\right)
\right)  ,
\]%
\begin{equation}
f_{x,t}^{-1}\left(  u_{t}^{\tau},dx^{\prime}\right)  =\left\{  v_{t}^{\tau
}:f_{x,t}^{\tau}\left(  u_{t}^{\tau},v_{t}^{\tau}\right)  \in dx^{\prime
}\right\}  ,\label{2.4}%
\end{equation}
satisfy the Chapman-Kolmogorov equation
\begin{equation}
\int\limits_{x^{\prime}\in X}\pi_{x,r}^{\tau}\left(  u_{r}^{\tau},dx^{\prime
}\right)  \pi_{x^{\prime},r+\tau}^{\tau^{\prime}}\left(  u_{r+\tau}%
^{\tau^{\prime}},dx^{\prime\prime}\right)  =\pi_{x,r}^{\tau+\tau^{\prime}%
}\left(  u_{r}^{\tau+\tau^{\prime}},dx^{\prime\prime}\right) \label{2.5}%
\end{equation}
for all $r\in\mathbb{R},$ $\tau,\tau^{\prime}>0$ and $u_{r}^{\tau+\tau
^{\prime}}=\left(  u_{r}^{\tau},u_{r+\tau}^{\tau^{\prime}}\right)  $, so that
the sufficient statistics form a controllable Markov process.
\end{theorem}

\begin{proof}
The existence of the sufficient statistics is determined by the \textit{a
posteriori} mapping according to the expression (\ref{1.8}), which gives in
correspondence with (\ref{2.1})
\begin{equation}
\varrho_{t}^{\tau}=\varrho\mathrm{M}_{\varrho,t}\left(  u_{t}^{\tau}%
,v_{t}^{\tau}\right)  =\varrho_{f_{x,t}^{\tau}\left(  u_{t}^{\tau},v_{t}%
^{\tau}\right)  ,t+\tau^{\prime}}\label{2.6}%
\end{equation}
for $\varrho=\varrho_{x,t}$, thus proving the first statement of Theorem
\ref{Theorem2}. The transition mappings $x\mapsto f\left(  u_{t}^{\tau}%
,v_{t}^{\tau}\right)  $ in correspondence with the Theorem \ref{Theorem1}
satisfy the semigroup property with respect to their composition
\begin{equation}
f_{r+\tau}^{\tau^{\prime}}\left(  u_{r+\tau}^{\tau^{\prime}},v_{r+\tau}%
^{\tau^{\prime}}\right)  \circ f_{r}^{\tau}\left(  u_{r}^{\tau},v_{r}^{\tau
}\right)  =f_{r}^{\tau+\tau^{\prime}}\left(  u_{r}^{\tau+\tau^{\prime}}%
,v_{r}^{\tau+\tau^{\prime}}\right)  .\label{2.7}%
\end{equation}
This yields, according to (\ref{1.5}), the equation
\begin{equation}
\int\limits_{x^{\prime}\in X}\Pi_{x,r}^{\tau}\left(  u_{r}^{\tau},dx^{\prime
}\right)  \Pi_{x^{\prime},r+\tau}^{\tau^{\prime}}\left(  u_{r}^{\tau^{\prime}%
},dx^{\prime\prime}\right)  =\Pi_{x,r}^{\tau+\tau^{\prime}}\left(  u_{r}%
^{\tau+\tau^{\prime}},dx^{\prime\prime}\right) \label{2.8}%
\end{equation}
in terms of the transfer-operator measures for the transitions $x\mapsto
dx^{\prime}$ specified in (\ref{2.3}) and (\ref{2.4}). \ For the states in the
class $\left\{  \varrho_{x,t}\right\}  $, (\ref{2.8}) is equivalent to the
Chapman-Kolmogorov equation (\ref{2.5}), as
\begin{equation}
\varrho_{x,t}\Pi_{x,t}^{\tau}\left(  u_{t}^{\tau},dx^{\prime}\right)
=\pi_{x,t}^{\tau}\left(  u_{t}^{\tau},dx^{\prime}\right)  \varrho_{x^{\prime
},t+\tau}\label{2.9}%
\end{equation}
\ in accordance with (\ref{2.1}). Thus equation (\ref{2.5}) determines a
Markov stochastic evolution $\widehat{x}\left(  t\right)  $ of the sufficient
coordinates $x\left(  t\right)  \in X$, which is described, according to the
main Kolmogorov theorem (see, e.g. \cite{Dav77}, p. 48 for a standard space
$X$), a Markov probability measure in the functional Borel space of the
trajectories $\left\{  x\left(  t\right)  \right\}  $. \ This completes the proof.
\end{proof}

We now discuss in more detail the special case, in which the transition
measures (\ref{1.9}) have the superoperator densities
\begin{equation}
\mathrm{P}_{t}^{\tau}\left(  u_{t}^{\tau},v_{t}^{\tau}\right)  Q=F_{t}^{\tau
}\left(  u_{t}^{\tau},v_{t}^{\tau}\right)  ^{\dagger}QF_{t}^{\tau}\left(
u_{t}^{\tau},v_{t}^{\tau}\right) \label{2.10}%
\end{equation}
with respect to a given consistent family of measures $\mu_{t}^{\tau}\left(
u_{t}^{\tau},dv_{t}^{\tau}\right)  $. \ Here the operators $F_{t}^{\tau
}\left(  u_{t}^{\tau},v_{t}^{\tau}\right)  $ are assumed to satisfy the
normalization condition
\begin{equation}
\int F_{t}^{\tau}\left(  u_{t}^{\tau},v_{t}^{\tau}\right)  ^{\dagger}%
F_{t}^{\tau}\left(  u_{t}^{\tau},v_{t}^{\tau}\right)  \mu_{t}^{\tau}\left(
u_{t}^{\tau},dv_{t}^{\tau}\right)  =I,\label{2.11}%
\end{equation}
in the Hilbert space $\mathcal{H}$, as well as the composition condition
\begin{equation}
F_{r+\tau}^{\tau^{\prime}}\left(  u_{r+\tau}^{\tau^{\prime}},v_{t+\tau}%
^{\tau^{\prime}}\right)  F_{r}^{\tau}\left(  u_{r}^{\tau},v_{r}^{\tau}\right)
=F_{r}^{\tau+\tau^{\prime}}\left(  u_{r}^{\tau+\tau^{\prime}},v_{r}^{\tau
+\tau^{\prime}}\right)  ,\label{2.12}%
\end{equation}
which guarantees the fulfillment of (\ref{1.13}).

One can easily see that the \textit{a posteriori} transfer operators
(\ref{1.10}) preserve the vectorial property of the states $\varrho_{\psi
}\left(  Q\right)  =\left\langle \psi|Q\psi\right\rangle $ such that
\[
\varrho_{\psi,t}^{\tau}\left(  Q\right)  =\left\langle \psi_{t}^{\tau}%
|Q\psi_{t}^{\tau}\right\rangle \equiv\varrho_{\psi_{t}^{\tau}}\left(
Q\right)  ,
\]
where $\psi_{t}^{\tau}=T_{\psi,t}^{\tau}\left(  u_{t}^{\tau},v_{t}^{\tau
}\right)  \psi$ for any unit vector $\psi\in\mathcal{H}$, $\tau>0$ and
\begin{equation}
T_{\psi,t}^{\tau}\left(  u_{t}^{\tau},v_{t}^{\tau}\right)  =\frac{F_{t}^{\tau
}\left(  u_{t}^{\tau},v_{t}^{\tau}\right)  }{\left\Vert F_{t}^{\tau}\left(
u_{t}^{\tau},v_{t}^{\tau}\right)  \psi\right\Vert }.\label{2.13}%
\end{equation}

\begin{corollary}
The normalized vectors $\psi\in\mathcal{H}$,$\left\Vert \psi\right\Vert =1 $
in the case (\ref{2.10}) form the space $X$ of sufficient coordinates $x=\psi$
specified by the a posteriori mappings $f_{\psi,t}^{\tau}\left(  u_{t}^{\tau
},v_{t}^{\tau}\right)  =T_{\psi,t}^{\tau}\left(  u_{t}^{\tau},v_{t}^{\tau
}\right)  \psi$ of the quasilinear form (\ref{2.13}) for the family of all
vectorial states $\varrho_{\psi}$.$_{{}}$
\end{corollary}

We note that the \textit{a posteriori} propagators $T_{\psi,t}^{\tau}\left(
u_{t}^{\tau},v_{t}^{\tau}\right)  $ satisfy the semigroup property
(\ref{2.7}):
\begin{equation}
T_{\psi^{\prime},r+\tau}^{\tau^{\prime}}\left(  u_{r+\tau}^{\tau^{\prime}%
},v_{r+\tau}^{\tau^{\prime}}\right)  T_{\psi,r}^{\tau}\left(  u_{r}^{\tau
},v_{r}^{\tau}\right)  =T_{\psi,r}^{\tau+\tau^{\prime}}\left(  u_{r}%
^{\tau+\tau^{\prime}},v_{r}^{\tau+\tau^{\prime}}\right)  ,\label{2.14}%
\end{equation}
where $\psi^{\prime}=T_{\psi,r}^{\tau}\left(  u_{r}^{\tau},v_{r}^{\tau
}\right)  \psi$. They are nonlinear (quasilinear), and in contrast with the
linear operators $F_{t}^{\tau}\left(  u_{t}^{\tau},v_{t}^{\tau}\right)  $ they
preserve the norm in $\mathcal{H}$. Only in the case discussed at the end of
Section \ref{Section3}, where $F_{t}^{\tau}\left(  u_{t}^{\tau},v_{t}^{\tau
}\right)  ^{\dagger}F_{t}^{\tau}\left(  u_{t}^{\tau},v_{t}^{\tau}\right)
=p_{t}^{\tau}\left(  u_{t}^{\tau},v_{t}^{\tau}\right)  I$, the operators
(\ref{2.13}) are $\psi$-independent isometries $T_{t}^{\tau}\left(
u_{t}^{\tau},v_{t}^{\tau}\right)  =F_{t}^{\tau}\left(  u_{t}^{\tau}%
,v_{t}^{\tau}\right)  /\sqrt{p_{t}^{\tau}\left(  u_{t}^{\tau},v_{t}^{\tau
}\right)  }$. \ We note, however, that the \textit{a priori} transfer
operators
\begin{equation}
\mathrm{M}_{t}^{\tau}\left(  u_{t}^{\tau}\right)  Q=\int F_{t}^{\tau}\left(
u_{t}^{\tau},v_{t}^{\tau}\right)  ^{\dagger}QF\left(  u_{t}^{\tau},v_{t}%
^{\tau}\right)  \mu_{t}^{\tau}\left(  u_{t}^{\tau},dv_{t}^{\tau}\right)
\label{2.15}%
\end{equation}
determining the controllable Markov dynamics of the quantum system
(\ref{1.9}), (\ref{2.10}) in the absence of observations, are not described by
the propagators $T_{t}^{\tau}\left(  u_{t}^{\tau}\right)  $, with the
exception of the degenerate case in which the \textit{a posteriori} states
coincide $\operatorname{mod}\mu_{t}^{\tau}$ with \textit{a priori} states,
i.e., actually do not depend on the results of the observations $v_{t}^{\tau
}.$

\section{\label{Section4}Optimal quantum feedback control}

Let us now discuss the optimal control of a quantum dynamical system with
observation $\left\{  \Pi_{t}^{\tau}\right\}  $. We assume that the
performance of the system is measured at each time $t$ by the mathematical
expectation $\left\langle \rho_{t},Q_{t}\left(  u_{t},dv_{t}\right)
\right\rangle $ of a certain physical quantity $Q_{t}\left(  u_{t}%
,dv_{t}\right)  \in\mathfrak{A}$ which continuously depends in strong operator
topology on the input state $u_{t}=\left\{  u\left(  t+\tau\right)  \right\}
_{\tau\geq0}$ and on the output event $dv_{t}=d\left\{  v\left(
t+\tau\right)  \right\}  _{\tau>0}$ according to the equation%
\begin{equation}
Q_{t}\left(  u_{t},dv_{t}\right)  =\Pi_{t}^{\tau}\left(  u_{t}^{\tau}%
,dv_{t}^{\tau}\right)  Q_{t+\tau}\left(  u_{t+\tau},dv_{t+\tau}\right)
+S_{t}^{\tau}\left(  u_{t}^{\tau},dv_{t}^{\tau}\right)  .\label{3.1}%
\end{equation}
Here $S_{t}^{\tau}\left(  u_{t}^{\tau},dv_{t}^{\tau}\right)  \in\mathfrak{A}$
are Hermitian operators having the integral form
\begin{equation}
S_{t}^{\tau}\left(  u_{t}^{\tau},dv_{t}^{\tau}\right)  =\int\limits_{0}^{\tau
}\Pi_{t}^{\tau^{\prime}}\left(  u_{t}^{\tau^{\prime}},dv_{t}^{\tau^{\prime}%
}\right)  S\left(  u\left(  t+\tau^{\prime}\right)  ,t+\tau^{\prime}\right)
d\tau^{\prime}\label{3.2}%
\end{equation}
for a Hermitian operator-function $S\left(  u,t\right)  =S\left(  u,t\right)
^{\dagger}$ completely determining (\ref{3.1}) for a certain boundary
condition $Q_{T}\left(  u_{T},dv_{T}\right)  =Q$ at the final time $T>t$. The
conditions for the existence of the integral \ref{3.2}, its continuous
dependence on $u_{t}^{\tau}$, and its $\sigma$-additivity with respect to
$dv_{t}^{\tau}$, requiring the continuity in $u\in U$ and measurability in
$t\in\mathbb{R}$ for the operator function $\left(  u,t\right)  \mapsto
S\left(  t,u\right)  \in\mathfrak{A}$ under strong operator topology, are
presumed to be fulfilled. The operator $Q$, specifying the terminal risk
$\left\langle \rho_{T},Q\right\rangle $, is assumed to be Hermitian-positive.

\begin{definition}
A measurable mapping $v_{t}\mapsto u_{t}\left(  v_{t}\right)  \in U_{t}$ is
called a non-anticipating control strategy if its components $u\left(
t+\tau,\cdot\right)  :v_{t}\mapsto u\left(  t+\tau\right)  $ are determined by
functions independent of $v_{t+\tau}$. It is called a retarded control
strategy if all $u\left(  t+\tau,\cdot\right)  $ are determined by functions
$v_{t}^{\tau^{\prime}}\mapsto u\left(  t+\tau,v_{t}^{\tau^{\prime}}\right)  $
for some measurable $\tau^{\prime}=\tau^{\prime}\left(  t+\tau\right)  <\tau$.
\ A non-anticipating strategy $u_{t}\left(  \cdot\right)  $ is called
admissible if the integral
\[
Q_{t}\left[  u\left(  \cdot\right)  \right]  =\int Q_{t}\left(  u_{t}\left(
v_{t}\right)  ,dv_{t}\right)
\]
exists in strong operator topology, and it is called optimal for an initial
state $\varrho_{t}=\varrho$ if it realizes the extremum
\begin{equation}
q\left(  \rho,t\right)  =\inf_{u_{t}\left(  \cdot\right)  \in U_{t}\left(
\cdot\right)  }\left\langle \rho,Q_{t}\left[  u_{t}\left(  \cdot\right)
\right]  \right\rangle ,\label{3.3}%
\end{equation}
where $U_{t}\left(  \cdot\right)  $ is a certain set of admissible strategies
$u_{t}\left(  \cdot\right)  $ $[\varepsilon$-optimal if $\left\langle
\rho,Q_{t}\left[  u_{t}\left(  \cdot\right)  \right]  \right\rangle $ exceeds
(\ref{3.3}) at most by $\varepsilon]$.
\end{definition}

We note that in accordance with (\ref{3.1}), a strategy $u_{t}\left(
\cdot\right)  $ is admissible with respect to $Q_{t}\left(  \cdot
,\cdot\right)  $ if and only if its segments $u_{t+\tau}\left(  \cdot\right)
$ are admissible strategies with respect to $Q_{t+\tau}\left(  \cdot
,\cdot\right)  $ for each fixed $v_{t}^{\tau}$, and if there exists measure
\begin{equation}
\Pi_{t}^{u,\tau}\left(  dv_{t}^{\tau}\right)  =\int\limits_{dv_{t}^{\tau}}%
\Pi_{t}^{\tau}\left(  u_{t}^{\tau}\left(  v_{t}^{\tau}\right)  ,dv_{t}^{\tau
}\right)  ,\label{3.4}%
\end{equation}
specifying the operator-valued integral
\begin{equation}
S_{t}^{\tau}\left[  u_{t}^{\tau}\left(  \cdot\right)  \right]  =\int
\limits_{0}^{\tau}\int\limits_{V_{t}^{\tau}}\Pi_{t}^{u,\tau^{\prime}}\left(
dv_{t}^{\tau^{\prime}}\right)  S\left(  t+\tau^{\prime},u\left(
t+\tau^{\prime},v_{t}^{\tau^{\prime}}\right)  \right) \label{3.5}%
\end{equation}
for each strategy segment $u_{t}^{\tau}\left(  \cdot\right)  $. The latter
holds for any delayed strategy that is admissible for a given boundary
condition $Q_{T}\left(  \cdot,\cdot\right)  =Q$.

\begin{theorem}
\label{Theorem3}Let the sets $U_{t}^{\tau}\left(  \cdot\right)  $ of
admissible strategy segments satisfy the condition
\begin{equation}
U_{t}^{\tau}\left(  \cdot\right)  \times U_{t+\tau}^{\tau^{\prime}}\left(
\cdot\right)  \subseteq U_{t}^{\tau+\tau^{\prime}}\left(  \cdot\right)  \text{
\ \ \ \ }\forall t\in\mathbb{R}\text{, \ \ \ \ }\tau,\tau^{\prime
}>0.\label{3.6}%
\end{equation}
Then the minimal risk (\ref{3.3}) as a function of the density operator $\rho$
and the time $t$ satisfies the functional equation
\begin{equation}
q\left(  \rho,t\right)  =\inf_{u_{t}^{\tau}\left(  \cdot\right)  \in
U_{t}^{\tau}\left(  \cdot\right)  }\left[  \left\langle \rho,S_{t}^{\tau
}\left[  u_{t}^{\tau}\left(  \cdot\right)  \right]  \right\rangle +\int
\pi_{\rho,t}^{u,\tau}\left(  dv_{t}^{\tau}\right)  q\left(  \rho\left(
v_{t}^{\tau}\right)  ,t+\tau\right)  \right]  ,\label{3.7}%
\end{equation}
where $\pi_{\rho,t}^{u,\tau}\left(  \cdot\right)  =\left\langle \rho,\Pi
_{t}^{u,\tau}\left(  \cdot\right)  I\right\rangle $, $\rho\left(  v_{t}^{\tau
}\right)  =\mathrm{M}_{\rho,t}^{\tau}\left(  u_{t}^{\tau}\left(  v_{t}^{\tau
}\right)  ,v_{t}^{\tau}\right)  _{\star}\rho$ denotes the probability measures
(\ref{1.7}) and \textit{a posteriori} states (\ref{1.8}) corresponding to an
admissible strategy $u=u_{t}^{\tau}\left(  \cdot\right)  $ and an initial
state $\varrho=\varrho_{t}$.
\end{theorem}

\begin{proof}
The proof of (\ref{3.7}) generalizes the proof of the Bellman equation
\cite{Bel57}. By substitution of (\ref{3.1}) into (\ref{3.3}) it reduces the
minimization over $u_{t}\left(  \cdot\right)  $ by the successive minimization
of (\ref{3.7}), first on $u_{t+\tau}\left(  \cdot\right)  $ and then on
$u_{t}^{\tau}\left(  \cdot\right)  $, which by condition (\ref{3.6}) yields
the same result as (\ref{3.3}). \ Since the integral (\ref{3.5}) does not
depend on $u_{t+\tau}\left(  \cdot\right)  $ and by definition,
\begin{equation}
\varrho\Pi_{t}^{u,\tau}\left(  dv_{t}^{\tau}\right)  =\pi_{\rho,t}^{u,\tau
}\left(  dv_{t}^{\tau}\right)  \varrho_{\rho,t}^{u,\tau},\label{3.8}%
\end{equation}
the first minimization entails finding the second term of the minimized sum
(\ref{3.7}):
\[
\inf_{u_{t+\tau}\left(  \cdot\right)  \in U_{t+\tau}\left(  \cdot\right)
}\int\pi_{\rho,t}^{u,\tau}\left(  dv_{t}^{\tau}\right)  \left\langle
\rho\left(  v_{t}^{\tau}\right)  ,\int Q_{t}\left(  u_{t}\left(  v_{t}\right)
,dv_{t}\right)  \right\rangle =\int\pi_{\rho,t}^{u,\tau}\left(  dv_{t}^{\tau
}\right)  q\left(  \rho\left(  v_{t}^{\tau}\right)  ,t+\tau\right)  .
\]

\end{proof}

In the case of a given boundary condition $q\left(  \rho,t\right)
=\left\langle \rho,Q\right\rangle $ the theorem proved above provides a
constructive method of synthesizing an optimal or $\varepsilon$-optimal
strategy $u_{\rho,t}^{T-t}\left(  v_{t}^{T-t}\right)  $ by the successive
minimization of (\ref{3.7}) in reverse time. \ In this case it is sufficient
to restrict the discussion to Markov admissible strategies described by
segments $u_{\rho,t^{\prime}}^{\tau}\left(  v_{t^{\prime}}^{\tau}\right)  $,
$\tau=T-t$, depending on the \textit{a priori} history $v_{t}^{\tau}$ only
through the agency of their dependence on the \textit{a posteriori} state
$\varrho=\varrho_{t}^{t^{\prime}-t}$ for any $t^{\prime}>t$. \ Accordingly,
the determination of the \textit{a posteriori} quantum states $\varrho
_{t}^{\tau}$, which generate the \textit{a posteriori} Markov process, enables
us to reduce the optimal quantum control problem to the classical problem of
stochastic control theory \cite{Str60}\cite{Bel57} with usual transition
probabilities and final risk functions
\[
s\left(  \rho,t,u\right)  =\left\langle \rho,S\left(  t,u\right)
\right\rangle ,\;q\left(  \rho,T\right)  =\left\langle \rho,Q\right\rangle ,
\]
determined by the operators of the corresponding quantum variables $S\left(
t,u\right)  $ and $Q$.

Let us consider the case in which the quantum states $\varrho$ are considered
in a certain class $\left\{  \varrho_{x,t}\right\}  $ for which sufficient
coordinates exist.

\begin{corollary}
Let $f_{t}^{\tau}:U_{t}^{\tau}\times V_{t}^{\tau}\mapsto X$ denote mappings
satisfying the conditions of Theorem \ref{Theorem2}. Then in problem
(\ref{3.3}) for $\varrho\in\left\{  \varrho_{x,t}\right\}  $ it is sufficient
to restrict the discussion to Markov strategies described by measurable
mappings $u_{t}^{\tau}:X\times V_{t}^{\tau}\mapsto U_{t}^{\tau} $ satisfying
the consistency condition
\begin{equation}
\left(  u_{x,r}^{\tau}\left(  v_{r}^{\tau}\right)  ,u_{x^{\prime},r+\tau
}^{\tau^{\prime}}\left(  v_{r+\tau}^{\tau^{\prime}}\right)  \right)
=u_{x,r}^{\tau+\tau^{\prime}}\left(  v_{r}^{\tau+\tau^{\prime}}\right)
,\label{3.9}%
\end{equation}
where $x^{\prime}=f_{x,t}^{u,\tau}\left(  v_{t}^{\tau}\right)  =f_{x,t}^{\tau
}\left(  u_{t}^{\tau}\left(  v_{t}^{\tau}\right)  ,v_{t}^{\tau}\right)  $.
\ In particular, the instantaneous control functions $u_{x}\left(
\tau\right)  $ for any $\tau\in\left[  t,T\right)  $ are determined by
functions $u\left(  \tau,x\right)  $ of the point state $x$ in accordance with
the equation
\begin{equation}
u_{x}\left(  t+\tau,v_{t}^{\tau}\right)  =u\left(  t+\tau,f_{x,t}^{u,\tau
}\left(  v_{t}^{\tau}\right)  \right)  .\label{3.10}%
\end{equation}

\end{corollary}

The foregoing assertion, which follows directly for the \textquotedblleft
maximum\textquotedblright\ sufficient coordinate $x\left(  t\right)
=\rho\left(  t\right)  $ from the optimality equation (\ref{3.7}), is readily
proved on the basis of the properties formulated for sufficient coordinates in
Theorem \ref{Theorem2}.

The further simplification of problem (\ref{3.3}) entails utilizing the
specific properties of the Markov process $x\left(  t\right)  =f_{x,0}%
^{u,t}\left(  v_{0}^{t}\right)  $, the role of which is logically assigned to
sufficient coordinates of the fewest possible dimensions.

For example, in the case where the generator for the transition probabilities
$\pi_{x,t}^{\tau}\left(  dx^{\prime}\right)  $, defined as the $t$-continuous
strong limit
\begin{equation}
L\left(  t\right)  q\left(  t\right)  \left(  x\right)  =\lim_{\tau\searrow
0}\int\limits_{X}\pi_{x,t}^{\tau}\left(  u_{t}^{\tau},dx^{\prime}\right)
\left(  q\left(  x^{\prime},t+\tau\right)  -q\left(  x,t+\tau\right)  \right)
,\label{3.11}%
\end{equation}
exists on some set $\mathcal{D}\left(  X\right)  $ of bounded and measurable
functions $x\mapsto q\left(  x,t\right)  $, depending continuously on $t$, the
optimality equation (\ref{3.3}) is written in the infinitesimal form
\begin{equation}
-\frac{\partial}{\partial t}q\left(  x,t\right)  =\inf_{u\in U}\left\{
s\left(  x,t,u\right)  +L\left(  t,u\right)  q\left(  t\right)  \left(
x\right)  \right\}  ,\label{3.12}%
\end{equation}
where $s\left(  x,t,u\right)  =\left\langle \rho_{xt},S\left(  t,u\right)
\right\rangle $. Equation (\ref{3.7}), which represents the standard Bellman
equation for controlled Markov processes in continuous time, can be used,
together with a boundary condition $q\left(  x,T\right)  =\left\langle
\rho_{x,T},Q\right\rangle \in\mathcal{D}\left(  X\right)  $, to seek optimal
or $\varepsilon$-optimal Markov\ control functions $u\left(  t\right)  $
directly as functions $u\left(  t,x\right)  $ of the instantaneous state $x$.

\section{Quantum control with discrete observation}

As an example here we consider the controlled dynamics of a simple quantum
system described between discrete measurement times $T=\left\{  t_{k}\right\}
$ by the Schr\H{o}dinger equation
\begin{equation}
i\hbar\frac{\partial}{\partial t}\psi\left(  t\right)  =H\left(  t,u\left(
t\right)  \right)  \psi\left(  t\right)  ,\ \ \ \ t\in T.\label{4.1}%
\end{equation}
Here $H\left(  t,u\right)  $ is the controlled Hamiltonian, i.e., a
self-adjoint operator in $\mathcal{H}$ with a dense domain of definition
$\mathcal{D}\subseteq\mathcal{H}$, written in the usual form
\[
H\left(  t,u\right)  =H_{0}\left(  t\right)  +\sum\limits_{i=1}^{m}%
u_{i}\left(  t\right)  H_{i}\left(  t\right)  ,
\]
where $u_{i}\left(  t\right)  \in\mathbb{R}$; $H_{i}\left(  t\right)  $ are
simple\footnote{In other words, having one-sided limits. For unbounded
self-adjoint operators $H_{i}\left(  t\right)  $, $i=0,...,m$, this means that
$H\left(  t,u\left(  t\right)  \right)  \psi$ is a simple function for any
$\psi\in\mathcal{D}.$} functions of $t$. \ Under the stated assumption there
exists a unique consistent family $\left\{  T_{t}^{\tau}\right\}  $ of unitary
propagators $T_{t}^{\tau}\left(  u_{t}^{\tau}\right)  $, representing for any
$\psi\left(  t-\tau\right)  =\varphi\in\mathcal{D}$, a solution of (\ref{4.1})
between adjacent measurement times $t_{k}<t_{k+1}$ in the form $\psi\left(
t\right)  =T_{t-\tau}^{\tau}\left(  u_{t-\tau}^{\tau}\right)  \psi,$
$t\in\left[  t_{k},t_{k+1}\right)  $, where $\underset{\tau\searrow0}{\lim
}\psi\left(  t\right)  =\psi.$

Let $E_{v,k}$ denote Hermitian projectors, which determine orthogonal
decompositions $I=\sum\limits_{v\in V_{k}}E_{v,k}$ of the unit operator in
$\mathcal{H}$ and specify measurements at times $t_{k}$ of quantum physical
quantities described by self-adjoint operators
\begin{equation}
A_{k}=\sum\limits_{v\in V_{k}}vE_{v,k}\label{4.2}%
\end{equation}
with discrete spectra $V_{k}\subseteq\mathbb{R}$.

As a result of measurement of the quantity $A_{k}$ there occurs a reduction
\cite{Neu55} of the quantum state, $\varrho\mapsto\varrho\Pi_{v,k},$ $v\in
V_{k}$, described by the superoperators $\Pi_{v,k}Q=E_{v,k}QE_{v,k},$ which
determines \textit{a priori} transfer operators
\begin{equation}
\Pi_{k}Q=\sum\limits_{v\in V_{k}}E_{v,k}QE_{v,k}.\label{4.3}%
\end{equation}

The states $\varrho_{v,k}=\varrho\Pi_{v,k}$ to which the system transfers
instantaneously depending on the result of this measurement $v\in V_{k}$ are
normalized to the probabilities $\pi_{v,k}=\left\langle \rho,E_{v,k}%
\right\rangle $ of these transitions, where if $\varrho=\varrho_{\psi}$ is a
vector state $\varrho_{\psi}\left(  Q\right)  =\left\langle \psi
|Q\psi\right\rangle $, the states $\varrho_{v,k}$ are also vectorial,
determined by the projections $\psi_{v,k}=E_{v,k}\psi$. The product
$E_{v,k}T_{t}^{\tau}\left(  u_{t}^{\tau}\right)  =F_{v,t}^{\tau}\left(
u_{t}^{\tau}\right)  $ for $\tau=t_{k}-t$ determines a transformation
$\psi\left(  t\right)  \mapsto E_{v,k}\psi\left(  t_{k}\right)  $
corresponding to the evolution (\ref{4.1}) on the interval $\left[
t,t_{k}\right)  $ with subsequent measurement of the quantity $A_{k}$.

We introduce the notation $F_{v,k}\left(  u_{k}\right)  =F_{v,t_{k}}^{\tau
_{k}}\left(  u_{t_{k}}^{\tau_{k}}\right)  $, where $\tau_{k}=t_{k+1}-t_{k}$,
and we set $V_{t}^{\tau}=\prod\limits_{k\in K_{t}^{\tau}}V_{k}$, where
$K_{t}^{\tau}=\left\{  k:t_{k}\in\left[  t,t+\tau\right)  \right\}  $ is the
set of all indices of times in the interval $\left(  t,t+\tau\right)  $ (in
the case of an empty set $K_{t}^{\tau}=\emptyset$ we assume that $V_{t}^{\tau
}$ consists of some single point $\left\{  w\right\}  $).

\begin{proposition}
Let the set $K_{t}^{s-t}$ be finite for any $t<s$. \ Then the chronological
product
\begin{equation}
F_{v,t}^{s-t}\left(  u_{t}^{s-t}\right)  =T_{t_{1}}^{s-t_{1}}\left(  u_{t_{1}%
}^{s-t_{1}}\right)  F_{v_{1},l-1}\left(  u_{l-1}\right)  \ldots F_{v_{k+1}%
,k}\left(  u_{k}\right)  F_{v_{k},t}^{t_{k}-t}\left(  u_{t}^{t_{k}-t}\right)
,\label{4.4}%
\end{equation}
where $k=\min K_{t}^{s-t}$, $l=\max K_{t}^{s-t}$, and $v=\left(  v_{k}%
,\ldots,v_{l}\right)  =v_{t}^{s-t}$, determines controllable quantum dynamical
system described by superoperators $\left\{  \Pi_{t}^{\tau}\right\}  $ of the
form (\ref{1.9}), (\ref{2.10}):
\begin{equation}
\Pi_{v,t}^{\tau}\left(  u_{t}^{\tau}\right)  Q=F_{v,t}^{\tau}\left(
u_{t}^{\tau}\right)  ^{\dagger}QF_{v,t}^{\tau}\left(  u_{t}^{\tau}\right)
,\label{4.5}%
\end{equation}
under the counting measure $\mu_{t}^{\tau}=1$ on $V_{t}^{\tau}\ni v.$
\end{proposition}

The proof is the verification of conditions (\ref{2.11}) and (\ref{2.12}),
which take the form
\begin{equation}
\sum\limits_{v\in V_{t}^{\tau}}F_{v,t}^{\tau}\left(  u_{t}^{\tau}\right)
^{\dagger}F_{v,t}^{\tau}\left(  u_{t}^{\tau}\right)  =I\text{ \ \ \ \ }\forall
u_{t}^{\tau}\in U_{t}^{\tau},\label{4.6}%
\end{equation}%
\begin{equation}
F_{v^{\prime},r+\tau}^{\tau^{\prime}}\left(  u_{r+\tau}^{\tau^{\prime}%
}\right)  F_{v,r}^{\tau}\left(  u_{r}^{\tau}\right)  =F_{\left(  v^{\prime
},v\right)  ,r}^{\tau^{\prime}+\tau}\left(  u_{r}^{\tau^{\prime}+\tau}\right)
,\label{4.7}%
\end{equation}
where $v^{\prime}\in V_{r+\tau}^{\tau^{\prime}}$, $v\in V_{r}^{\tau}$. They
are easily verified by induction, owing to the finiteness of the product (4.4).

Because of the spatial form (\ref{4.5}) of the consistent family $\left\{
\Pi_{v,t}\right\}  $, on the basis of Corollary 1 we infer that the space $X$
of normalized vectors $\psi\in\mathcal{H}$, $\left\Vert \psi\right\Vert =1 $
forms a space of sufficient coordinates, the \textit{a posteriori} evolution
$\psi\mapsto T_{v,t}^{\tau}\left(  u_{t}^{\tau},v_{t}^{\tau}\right)  \psi$ of
which is described by the nonlinear propagators (\ref{2.13}): $T_{v,t}^{\tau
}\left(  u_{t}^{\tau}\right)  /\left\Vert T_{v,t}^{\tau}\left(  u_{t}^{\tau
}\right)  \psi\right\Vert $, and the \textit{a priori} evolution by transfer
operators of the form (\ref{2.15}):
\[
\Pi_{t}^{\tau}\left(  u_{t}^{\tau}\right)  Q=\sum\limits_{v\in V_{t}^{\tau}%
}F_{v,t}^{\tau}\left(  u_{t}^{\tau}\right)  ^{\dagger}QF_{v,t}^{\tau}\left(
u_{t}^{\tau}\right)  .
\]

We give special consideration to the case of complete measurements described
by the operators $A_{k}$ with a non-degenerate spectrum.

\begin{proposition}
Let $\left\{  \psi_{v,k}\right\}  _{v\in V_{k}}$ denote the complete
orthonormal systems of eigenvectors of the operators $A_{k}$, and let
$E_{v,k}$ be the corresponding one-dimensional projectors onto $\psi_{v,k}$.
\ Then the \textit{a posteriori} states at the times $\left\{  t_{k}\right\}
$ are vector states, which are completely determined by the last result of
measurement $v_{k}\in V_{k}:$%
\begin{equation}
\left\langle \rho_{t}^{t_{k}-t},Q\right\rangle =\left\langle Q\psi_{v_{k}%
,k},\psi_{v_{k},k}\right\rangle \text{ \ \ \ \ }\forall t<t_{k},\text{
\ \ \ \ }\varrho=\varrho_{t}\text{,}\label{4.8}%
\end{equation}
and the measurement process $\left\{  v_{k}\right\}  $ is a Markov process,
which is described by the controllable transition probabilities
\begin{equation}
\pi_{v,k}\left(  u_{k},v_{k}\right)  =\left\vert \left\langle \psi_{v_{k}%
,k+1}|T_{k}\left(  u_{k}\right)  \psi_{v_{k},k}\right\rangle \right\vert
{{}^2}%
,\label{4.9}%
\end{equation}
where $T_{k}\left(  u_{k}\right)  =T_{t_{k}}^{\tau_{k}}\left(  u_{t_{k}}%
^{\tau_{k}}\right)  ,$ $\tau_{k}=t_{k+1}-t_{k}.$
\end{proposition}

This proposition follows from the property
\begin{equation}
E_{v,k}QE_{v,k}=\left\langle \psi_{v,k}|Q\psi_{v,k}\right\rangle E_{v,k}%
\end{equation}
of the one-dimensional orthogonal projection operators $E_{v,k}$ corresponding
to the eigenvectors $\psi_{v,k}$, so that the application of any state
$\varrho$ to (\ref{4.5}) at $t=t_{k}-\tau$ leads to (\ref{4.8}), up to
normalization. \ Since the \textit{a posteriori} state (\ref{4.8}) does not
depend on the previous measurements, the conditional probability given by
expression (\ref{4.9}) for the event $v_{k+1}=v$ and fixed preceding results
is Markovian.

In the proposition proved above, the controllable sufficient coordinate
$x_{k}=v_{k}$ can be used, provided only that the quantum system is analyzed
at discrete measurement times $\left\{  t_{k}\right\}  .$

We now consider the optimal control problem for a discretely observed quantum
system. \ Let the control performance, as a function of the initial $t,$ be
described by an operator (\ref{3.1}), which is determined by the integral
(\ref{3.2}) of some operator-valued function $S\left(  t,u\right)
:\mathcal{H}\mapsto\mathcal{H}$.

\begin{proposition}
Under the conditions of Theorem \ref{Theorem3}, for a vector initial state
$\varrho_{\psi}$ the minimal risk
\begin{equation}
q\left(  \psi,t\right)  =\inf_{u_{t}\left(  \cdot\right)  \in U_{t}\left(
\cdot\right)  }\int\limits_{v_{t}}\left\langle \psi|Q_{t}\left(  u_{t}\left(
v_{t}\right)  ,dv_{t}\right)  \psi\right\rangle \label{4.10}%
\end{equation}
in the intervals $\left(  t_{k},t_{k+1}\right)  $ between measurements
satisfies the functional equation in variational derivatives
\begin{equation}
-\frac{\partial}{\partial t}q\left(  \psi,t\right)  =\inf_{u\in U\left(
t\right)  }\left(  \left\Vert \psi\right\Vert _{S\left(  t,u\right)  }%
^{2}+2\hbar^{-1}\operatorname{Im}\left\langle \delta q\left(  \psi,t\right)
/\delta\psi|H\left(  t,u\right)  \psi\right\rangle \right)  ,\label{4.11}%
\end{equation}
where $\left\Vert \psi\right\Vert _{S}^{2}=\left\langle \psi|S\psi
\right\rangle $. At the measurement time instances $\left\{  t_{k}\right\}  $
it satisfies the recursive equation
\begin{equation}
q_{k}\left(  \psi\right)  =\inf_{u_{k}\in U_{k}}\left(  \left\Vert
\psi\right\Vert _{S_{k}\left(  u_{k}\right)  }^{2}+\sum\limits_{v\in V_{k+1}%
}\pi_{v,k}^{u}\left(  \psi\right)  q_{k+1}\left(  \psi\left(  v\right)
\right)  \right)  ,\label{4.12}%
\end{equation}
which determines the boundary values $q\left(  t_{k}-0,\psi\right)
=q_{k}\left(  \psi\right)  $ for (\ref{4.11}). \ Here $\pi_{v,k}^{u}\left(
\psi\right)  =\left\Vert T_{v,k}\left(  u_{k}\right)  \psi\right\Vert ^{2}$,
$\psi\left(  v\right)  =T_{v,k}\left(  u_{k}\right)  \psi/\sqrt{\pi_{v,k}%
^{u}\left(  \psi\right)  }$, and
\begin{equation}
S_{k}\left(  u_{k}\right)  =\int\limits_{0}^{\tau}T_{t_{k}}^{t-t_{k}}\left(
u_{t_{k}}^{t-t_{k}}\right)  ^{\dagger}S\left(  t,u\left(  t\right)  \right)
T_{t_{k}}^{t-t_{k}}\left(  u_{t_{k}}^{t-t_{k}}\right)  dt.\label{4.13}%
\end{equation}

\end{proposition}

Equation (\ref{4.11}) is readily proved on the assumption of analyticity of
the function $\psi\mapsto q\left(  \psi,t\right)  $ which is natural for a
quadratic boundary condition $q\left(  \psi,t\right)  =\left\Vert
\psi\right\Vert _{Q}^{2}$ at some final time $T$. \ Here (\ref{4.11})
represents a functional version of the Bellman equation corresponding to the
Schr\H{o}dinger equation (\ref{4.1}) and a quadratic transition cost function
$S\left(  t,u,\psi\right)  =\left\Vert \psi\right\Vert _{S\left(  t,u\right)
}^{2}$. \ Equation (\ref{4.12}) follows directly from (\ref{3.7}) for
$t=t_{k},$ $\tau=t_{k+1}-t_{k}$ and $\varrho=\varrho_{\psi}$ if it is taken
into account that the integral (\ref{3.2}) now has the form (\ref{4.13}).

In conclusion we consider the optimal control problem described above in the
complete measurement case. \ Making use of the fact that the process of
complete measurement at discrete times $\left\{  t_{k}\right\}  $ induces a
Markov sufficient coordinate $x_{k}=v_{k}$, from (\ref{4.12}) we deduce the
customary equation
\begin{equation}
q_{k}\left(  v_{k}\right)  =\inf_{u_{k}\in U_{k}}\left(  s_{k}\left(
u_{k},v_{k}\right)  +\sum\limits_{v\in V_{k+1}}\pi_{v,k}\left(  u_{k}%
,v_{k}\right)  q_{k+1}\left(  v\right)  \right)  ,\label{4.14}%
\end{equation}
which describes the optimum risk for the control of a discrete Markov process
$\left\{  v_{k}\right\}  $ with the transition probabilities (\ref{4.9}), a
cost function $s_{k}\left(  u_{k},v_{k}\right)  =\left\Vert \psi_{v_{k}%
}\right\Vert _{S_{k}\left(  u_{k}\right)  }^{2}$ and a boundary condition of
the form $q_{k}\left(  v\right)  =\left\Vert \psi_{v_{k}}\right\Vert _{Q}^{2}%
$. \ The solution of derived Bellman equation (\ref{4.14}) can be easily
modelled on a computer by standard dynamic programming methods for the
piecewise-constant admissible strategies, for which $U_{k}=U\left(
t_{k}\right)  \subseteq\mathbb{R}^{m}.$

\appendix

\section{Notations, Definitions and Facts}

\begin{enumerate}
\item Let $\left\{  Q_{i}\right\}  _{i\in\mathrm{I}}$ be a family of
self-adjoint operators acting in a complex Hilbert space $\mathcal{H}$. \ The
von Neumann algebra generated by the family $\left\{  Q_{i}\right\}  $ is
defined as the minimal weakly closed self-adjoint sub-algebra $\mathfrak{A}$
of bounded operators in $\mathcal{H}$ containing the spectral projectors of
this operators, along with the unit operator $I$. \ It consists of all bounded
operators that commute with the commutant $\left\{  Q_{i}\right\}  ^{\prime
}=\left\{  Q:QQ_{i}=Q_{i}Q\text{ \ \ }\forall i\in\mathrm{I}\right\}  $, i.e.,
it is the second commutant $\mathfrak{A}=\left\{  Q_{i}\right\}
^{\prime\prime}$ of the family $\left\{  Q_{i}\right\}  $. \ The latter can be
taken as the definition of the von Neumann algebra generated by the family
$\left\{  Q_{i}\right\}  $ in the case of unbounded self-adjoint operators
$Q_{i}$ densely defined on a domain $\mathcal{D}\subseteq\mathcal{H}$. \ The
simplest example of von Neumann algebra is the algebra $\mathcal{B}\left(
\mathcal{H}\right)  $ of all bounded operators acting in $\mathcal{H}$
\cite{Dix69}.

\item A state on a von Neumann algebra $\mathfrak{A}$ is defined as a linear
ultraweakly continuous functional $\varrho:\mathfrak{A}\rightarrow\mathbb{C}$
(which will be denoted as $\varrho\left(  Q\right)  =\left\langle
\rho,Q\right\rangle $) satisfying the positivity and normalization conditions
\begin{equation}
\left\langle \rho,Q\right\rangle \geq0,\text{ \ \ \ \ }\forall Q\geqq0\text{,
\ \ \ \ }\left\langle \rho,I\right\rangle =1\label{A.1}%
\end{equation}
[$Q\geqq0$ signifies the nonnegative definiteness $\left\langle \psi
|Q\psi\right\rangle \geqq0$ $\forall\psi\in\mathcal{H}$ called Hermitian
positivity of $Q$]. It is described by the density operators $\rho$ as the
elements of the algebra $\mathfrak{A}$ with respect to a standard pairing
$\left\langle \rho,Q\right\rangle $.\ The linear span of all\emph{\ }states on
$\mathfrak{A}$ is a Banach subspace $\mathfrak{A}_{\star}$ of the dual space
$\mathfrak{A}^{\star}$, called predual to $\mathfrak{A}$ as $\mathfrak{A}%
_{\star}^{\star}=\mathfrak{A}$.\ A state $\varrho$ is called vector state if
$\left\langle \rho,Q\right\rangle =\left\langle \psi|Q\psi\right\rangle $
($\varrho=\varrho_{\psi}$) for some $\psi\in\mathcal{H}$. \ Any state is a
closed convex hull of vector states $\varrho_{\psi}$, $\left\Vert
\psi\right\Vert =1$. \ If on an algebra $\mathfrak{A}$ there exists a normal
semi-finite trace $Q\mapsto\mathrm{tr}\left\{  Q\right\}  $, then the states
on $\mathfrak{A}$ can be described by the density operators $\rho
\in\mathfrak{A}$ (or affiliated to $\mathfrak{A}$, if they are unbounded),
determining $\varrho$ by means of the bilinear form $\left\langle
\rho,Q\right\rangle =\mathrm{tr}\left\{  \rho Q\right\}  $. \ For the case
$\mathfrak{A}=\mathcal{B}\left(  \mathcal{H}\right)  $ the density operator
$\rho$ is any nuclear positive operator with unit trace \cite{Dix69}.

\item Let $\mathfrak{A}_{1}$, $\mathfrak{A}_{2}$ be von Neumann algebras in
respective Hilbert spaces $\mathcal{H}_{1}$ and $\mathcal{H}_{2}$, and let
$\mathrm{M}:\mathfrak{A}_{2}\rightarrow\mathfrak{A}_{1}$ be a linear map that
transforms the operators $Q_{2}\in\mathfrak{A}_{2}$ into operators $Q_{1}%
\in\mathfrak{A}_{1}$ (superoperator, in the terminology of \cite{Ing75}). \ We
shall call the operator $\mathrm{M}$ a transfer operator if it is ultraweakly
continuous, completely positive in the sense
\begin{equation}
\sum\limits_{i,k=1}\left\langle \psi_{i}|\mathrm{M}\left(  Q_{i}^{\dagger
}Q_{k}\right)  \psi_{k}\right\rangle \geqq0\text{, \ \ \ \ }\forall Q_{i}%
\in\mathfrak{A}_{2},\text{ \ \ \ \ }\psi_{i}\in\mathcal{H}_{1},\label{A.2}%
\end{equation}
($i=1,\ldots,n<\infty$), and unity-preserving: $\mathrm{M}I_{2}=I_{1}$. \ In
this case the composition $\varrho_{1}\mathrm{M}$ with a state $\varrho
_{1}:\mathfrak{A}_{1}\rightarrow\mathbb{C}$ is a (normal) state $\mathfrak{A}%
_{2}\rightarrow\mathbb{C}$ described by the predual action of the
superoperator $\mathrm{M}$ on $\rho_{1}$:
\[
\left\langle \rho_{1},\mathrm{M}Q_{2}\right\rangle =\left\langle
\mathrm{M}_{\star}\rho_{1},Q_{2}\right\rangle ,\forall Q_{2}\in\mathfrak{A}%
_{2},\rho_{1}\in\mathfrak{A}_{1}^{\backprime}.
\]
\ A transfer operator $\mathrm{M}$ is called spatial if
\begin{equation}
\mathrm{M}Q_{2}=T^{\dagger}Q_{2}T\text{ \ \ \ or \ \ \ }\mathrm{M}_{\star}%
\rho_{1}=T\rho_{1}T^{\dagger},\label{A.3}%
\end{equation}
where $T:\mathcal{H}_{1}\rightarrow\mathcal{H}_{2}$ is a linear isometric
operator, $T^{\dagger}T=I$, called the propagator, and $T^{\dagger}$ is the
adjoint operator. \ Every transfer operator on $\mathfrak{A}_{2}%
=\mathcal{B}\left(  \mathcal{H}_{2}\right)  $ has a decomposition as a closed
convex hull of spatial transfer operators.

\item Let $V$ be a measurable space, and $\mathcal{B}$ its Borel $\sigma
$-algebra. \ A mapping $\Pi:dv\in\mathcal{B}\mapsto\Pi\left(  dv\right)  $
with values $\Pi\left(  dv\right)  $ in ultraweakly continuous, completely
positive superoperators $\mathfrak{A}_{2}\rightarrow\mathfrak{A}_{1}$ is
called a transfer-operator measure if for any $\rho_{1}\in\mathfrak{A}_{1}$,
$Q_{2}\in\mathfrak{A}_{2}$ the numerical function
\[
\left\langle \Pi\left(  dv\right)  _{\star}\rho_{1},Q_{2}\right\rangle
=\left\langle \rho_{1},\Pi\left(  dv\right)  Q_{2}\right\rangle
\]
of the set $dv\subseteq V$ is a countably additive measure normalized to unity
for $Q_{2}=I$. \ In other words, $\Pi\left(  dv\right)  $ is an
operator-valued measure that is $\sigma$-additive in the weak (strong)
operator sense and for $dv=V$ is equal to some transfer operator $\mathrm{M}$.
\ The quantum-state transformations $\rho_{1}\mapsto\rho_{2}$ corresponding to
ideal measurements are described by transfer-operator measures of the form
\begin{equation}
\Pi\left(  dv\right)  Q=F\left(  v\right)  ^{\dagger}QF\left(  v\right)
\mu\left(  dv\right)  ,\label{A.4}%
\end{equation}
where $F\left(  v\right)  $ denotes linear operators $\mathcal{H}%
_{1}\rightarrow\mathcal{H}_{2}$, the integral under a positive numerical
measure $\mu$ on $V$ is interpreted in strong operator topology, and $\int
F^{\dagger}\left(  v\right)  F\left(  v\right)  \mu\left(  dv\right)  =I_{1}.$
\ Every transfer-operator $\mathrm{M}:\mathfrak{A}_{2}\rightarrow
\mathfrak{A}_{1}$ for $\mathfrak{A}_{2}=\mathcal{B}\left(  \mathcal{H}\right)
$ can be represented by the integral (\ref{A.4}) with respect to $dv\subseteq
V$ of some ideal measure $\Pi\left(  dv\right)  $.
\end{enumerate}


\begin{thebibliography}{10}

\bibitem{Be74}
V.~P. Belavkin.
\newblock Optimal linear random filtration of quantum boson signals.
\newblock {\em Problems of Control and Information Theory}, 3:47--62, 1974.

\bibitem{Be78a}
V.~P. Belavkin.
\newblock Operational theory of quantum stochastic processes.
\newblock In {\em Proc of {VIIth} Conference on Coding Theory and Information
  Transmission}, volume~1, pages 23--28, Moscow--Vilnius, 1978.

\bibitem{Be78}
V.~P. Belavkin.
\newblock Optimal quantum filtration of markovian signals.
\newblock {\em Problems of Control and Information Theory}, 7(5):345--360,
  1978.

\bibitem{Be78b}
V.~P. Belavkin.
\newblock Quantum statistical theory of control.
\newblock In {\em Proc. of U S S R Conference on the Control of Laser
  Radiation}, volume~2, pages 19--23, Tashkent, 1978.

\bibitem{Be79}
V.~P. Belavkin.
\newblock Optimal measurement and control in quantum dynamical systems.
\newblock Technical Report 411, Instytut Fizyki, Copernicus University, Torun',
  1979.

\bibitem{Be80a}
V.~P. Belavkin.
\newblock Optimization of quantum observation and control.
\newblock In {\em Notes in Control and Inform Sci, Proc of 9th {IFIP} Conf on
  Optimizat Techn, Warszawa 1979}, volume~1, pages 141--149. Springer-Verlag,
  1980.

\bibitem{Bel57}
R.~Bellman.
\newblock {\em Dynamic Programming}.
\newblock Princeton Univ. Press, Princeton, New Jersey, 1957.

\bibitem{But_Sam80}
A.~G. Butkovski and Y.~I. Samoilenko.
\newblock Controlability of quantum systems.
\newblock {\em Dokl. Akad. Nauk SSSR}, 250(1):51--55, 1980.

\bibitem{But_Sam79}
A.~G. Butkovskii and Y.~I. Samoilenko.
\newblock Control of quantum systems.
\newblock {\em Avtomat. Telemekh.}, pages 5--25, 1979.

\bibitem{Dav77}
E.~V. Davies.
\newblock Quantum communication systems.
\newblock {\em IEEE Trans. Inf. Theory}, IT-23:530--534, 1977.

\bibitem{Dix69}
J.~Dixmer.
\newblock {\em Les Algebres {D'Operateurs} Dans {L'Espace} Hilbertien}.
\newblock Gauthier-Villars, Paris, 1969.

\bibitem{Emc72}
G.~G. Emch.
\newblock {\em Algebraic Methods in Statistical Mechanics and Quantum Field
  Theory}.
\newblock Wiley-Interscience, New York, 1972.

\bibitem{Ing75}
E.~S. Ingarden.
\newblock Quantum information theory.
\newblock In {\em Preprint No. 317, Inst. Phys. {UMK}, Torum}, 1975.

\bibitem{Kol74}
A.~N. Kolmogorov.
\newblock {\em Fundamental Concepts of Probability Theory [in Russian]}.
\newblock Nauka, Moscow, 1974.

\bibitem{Kra73}
A.~A. Krasovskii.
\newblock Ultimate precision of microcontrol.
\newblock {\em Avtomat. Telemekh}, 12:27--39, 1973.

\bibitem{Kra74}
A.~A. Krasovskii.
\newblock Ultimate precision of microobservation and microcontrol.
\newblock {\em Tekh. Kibern.}, 3:177--187, 1974.

\bibitem{Lan_Lif63}
L.~D. Landau and E.~M. Lifshits.
\newblock {\em Principles of Quantum Mechanics [in Russian]}.
\newblock Fizmatgiz, Moscow, 1963.

\bibitem{Neu55}
J.~V. Neumann.
\newblock {\em Mathematical Foundations of Quantum Mechanics}.
\newblock Princeton Univ. Press, Princeton, New Jersey, 1955.

\bibitem{Sam72}
Y.~I. Samoilenko.
\newblock Electromagnetic control of charged particles with allowance for
  stochastic and quantum effects.
\newblock {\em Controllable Stochastic Processes and Systems [in Russian]},
  pages 120--140, 1972.

\bibitem{Str60}
R.~L. Stratonovich.
\newblock Conditional markov processes.
\newblock {\em Teor. Veroyatn. Primen.}, pages 172--195, 1960.

\end{thebibliography}

\end{document}